\documentclass[12pt]{article}
\usepackage{epsfig}

\newcommand{\be}{\begin{eqnarray}}
\newcommand{\ee}{\end{eqnarray}}
\newcommand{\sll}{\raise.15ex\hbox{$/$}\kern-.43em\hbox{$l$}}
\newcommand{\slp}{\raise.15ex\hbox{$/$}\kern-.43em\hbox{$p$}}
\newcommand{\slq}{\raise.15ex\hbox{$/$}\kern-.43em\hbox{$q$}}
\newcommand{\slk}{\raise.15ex\hbox{$/$}\kern-.43em\hbox{$k$}}
\newcommand{\slepsilon}{\raise.15ex\hbox{$/$}\kern-.53em\hbox{$\epsilon$}}
\newcommand{\gsim}{\mbox{\raisebox{-0.6ex}{$\stackrel{>}{\sim}$}}\:}
\newcommand{\lsim}{\mbox{\raisebox{-0.6ex}{$\stackrel{<}{\sim}$}}\:}

\begin{document}

\bibliographystyle{unsrt}
\footskip 1.0cm

\thispagestyle{empty}
\begin{flushright}
INT--PUB 05--30
\end{flushright}
\vspace{0.1in}

\begin{center}{\Large \bf {Geometric scaling violations in the
  central rapidity region of d+Au collisions at RHIC}}\\

\vspace{1in}
{\large  Adrian Dumitru$^a$, Arata Hayashigaki$^a$ and 
  Jamal Jalilian-Marian$^b$}\\

\vspace{.2in}
{\it 
$^a$Institut f\"ur Theoretische Physik, J.~W.~Goethe Universit\"at\\
Max-von-Laue Strasse 1\\
D-60438 Frankfurt am Main, Germany\\
$^b$Institute for Nuclear Theory, University of Washington,
Seattle, WA 98195 }

\end{center}

\vspace*{25mm}

\begin{abstract}

\noindent 
We show that geometric scaling is satisfied to good accuracy in the
forward region of d+Au collisions at RHIC. Scaling violations do show
up, however, at mid-rapidity, and the anomalous dimension of the
small-$x$ gluon distribution evolves to near its DGLAP limit
for transverse momenta of a few GeV. This represents a first
consistency check of RHIC deuteron-nucleus and HERA DIS
phenomenology, and of the universality of the underlying Color Glass
Condensate (CGC) theory, which describes both phenomena. It also
reconciles successful leading-twist LO and NLO perturbative QCD
computations of mid-rapidity particle production with small-$x$
evolution.  Finally, we introduce a new parameterization for the
anomalous dimension of the small-$x$ gluon distribution which properly
reproduces known theoretical limits at large rapidity, at large
virtuality, and on the saturation boundary, and still fits the
available data from d+Au collisions at RHIC. We find indications that
sub-asymptotic terms in the rapidity-evolution of the anomalous
dimension are large.
\end{abstract}
\newpage

\section{Introduction}
The weak-coupling gluon saturation formalism~\cite{nonlin,JK} has been
used at HERA, rather successfully, in order to describe the $x$ and $Q^2$ 
dependence of the inclusive proton structure function $F_2(x,Q^2)$. 
Specifically, the
simple Golec-Biernat and W\"usthoff (GBW) model~\cite{GBW} provided a
very efficient qualitative ``summary'' of the small-$x$, moderate
$Q^2$ data in terms of an initial condition $Q_s^2(x_0)=Q_0^2$ for the
saturation momentum and its (constant) growth rate $\lambda =
\partial\log Q_s^2/\partial\log 1/x$. It led to the discovery of the
interesting phenomenon of ``geometric scaling''~\cite{Stasto:2000er},
implying that in the relevant range of $x$ and $Q^2$, the structure
function depends on $x$ and $Q^2$ exclusively via the scaling variable
$Q^2/Q_s^2(x)$. In other words, that the scattering cross section for
a dipole of size $r_t$ does not depend on $r_t$ and $x$ separately but
only through the combination $\rho = r_t Q_s(x)$.

This phenomenon does not find a natural explanation within the linear
Dokshitzer, Gribov, Lipatov, Altarelli, Parisi (DGLAP)~\cite{dglap}
and Balitsky, Fadin, Kuraev, Lipatov (BFKL)~\cite{bfkl} QCD evolution
equations in the absence of saturation boundary conditions. It has
been shown to arise from the BFKL equation {\em with} saturation
boundary conditions~\cite{ScalingViol}, in an expansion of the LO-BFKL
solution to first order about the saturation line, and in mean-field
approximation. That equation, with the corresponding boundary
condition, represents an approximation to the full non-linear ``Color
Glass Condensate'' (CGC) theory of QCD evolution. It is valid only on
one side of the saturation line $Q_s(x)$, in the dilute regime.

Moreover, refs.~\cite{ScalingViol} also computed the {\em scaling
violations}\footnote{We shall use this terminology throughout the
manuscript to refer to violations of {\em geometric scaling} away from
the saturation line, not to the violation of $Q^2$ scaling due to
DGLAP evolution.} which emerge away from the saturation boundary at
$Q^2\gg Q_s^2(x)$. They arise from the diffusion term in the expansion
of the LO-BFKL solution to second order about the saturation
saddle-point, which shifts the anomalous dimension $\gamma$ of the
target gluon distribution by $\Delta\gamma\sim
y^{-1}\,\log(1/r_tQ_s(y))$ as one moves away from the saturation
boundary ($y=\log 1/x$ denotes rapidity). By $Q_{gs}^2(x)\sim
Q_s^4(x)/ \Lambda^2$, scaling violations grow to order one ($\Lambda$
is a non-perturbative scale in the infrared, of order $\Lambda_{\rm
QCD}$); $Q_{gs}$ is the upper boundary of the so-called geometric
scaling window. Beyond this virtuality, the target gluon distribution
is described for example by the DLA form, if $y$ and $\log Q^2$ are
sufficiently large.

A fit to the HERA data in the (``extended'') geometric scaling window
$Q_s^2(x)<Q^2<Q_{gs}^2(x)$ has been performed by Iancu, Itakura and
Munier (IIM) in ref.~\cite{IIM} (see also~\cite{Forshaw}), employing a
dipole parameterization which represents an approximate solution to
the JIMWLK equations~\cite{jimwlk} governing the mean-field evolution
of color dipoles.  They find evidence for scaling violations in the
HERA data, as predicted by the CGC theory: a substantial increase of
$\gamma$ with $1/r_t$ is required to improve the agreement with the
data as compared to the original GBW dipole model.

In recent years, new data on single-inclusive hadron production at
semi-hard transverse momenta emerged from the Relativistic Heavy-Ion
Collider (RHIC). Here, we are interested in particular in data from
deuteron-gold collisions at central and forward rapidity (toward the
fragmentation region of the deuteron
projectile)~\cite{Arsene:2004ux,greg,STAR_y0,PHENIX_y0}, which may
probe high gluon densities in the gold target, and at the same time,
should be less affected by final-state effects than nucleus-nucleus
collisions.

A first semi-quantitative analysis of the d+Au data from RHIC was
performed by Kharzeev, Kovchegov and Tuchin (KKT) in ref.~\cite{dima},
identifying qualitative features of the expected CGC evolution and
providing an alternative dipole parameterization: the KKT model, to be
described below, was constructed to fit RHIC instead of HERA data.  In
ref.~\cite{aaj}, we extended the standard CGC approach to particle
production to include recoil effects of the large-$x$ projectile
partons, and found very good quantitative agreement with RHIC data at
large rapidity using either the KKT or the IIM dipole. It turned out
that a crucial feature of both dipole models lies in the fact that the
target anomalous dimension at large rapidity and moderate $p_t$ is
essentially constant, $\gamma\approx0.6$, and substantially smaller
than in the DGLAP regime ($\gamma_{DGLAP}\sim1$). Hence, for transverse
momenta of a few GeV and rapidity $y_h\gsim3$, the data points are in
either the saturation or extended scaling regions, where scaling
violations are essentially {\em absent} at RHIC energy.

In the present paper, we extend our previous calculation of forward
hadron production~\cite{aaj} to the
mid-rapidity region. Our main objective is to analyze whether
scaling violations, as predicted by the CGC theory and previous dipole
fits to HERA data, show up. This is a necessary condition for establishing
the CGC as a universal theory of QCD evolution near a saturation
boundary within the weak-coupling (semi-hard) regime.

The paper is organized as follows. In the next section, we introduce
the formula for the single-inclusive hadron production cross section
in high-energy proton (or deuteron) collisions with a heavy nucleus
which accounts for both small-$x$ evolution of the dense target and
full DGLAP evolution of the dilute projectile. We also present the KKT
dipole parameterization there. In section~\ref{sec_ScalViol} we apply
this formalism to hadron production at mid-rapidity and compare to
data from RHIC to check for the presence of scaling violations. In
section~\ref{sec_NewDipole} we modify the IIM dipole parameterization,
in particular that of the anomalous dimension, to fit the available
data from RHIC at central and at forward rapidity, which is done in
section~\ref{sec_applic}.  Finally, section~\ref{sec_summary} contains
our summary and conclusion.

\section{Proton-Nucleus Collisions within the CGC approach}
\label{sec2}
The single-inclusive hadron production cross section in p+A collision 
is given by~\cite{aaj}
\begin{eqnarray}
x_F{d\sigma^{p A \rightarrow h X} \over dx_F \, d^2 p_t \, d^2 b} &=& 
{1 \over (2\pi)^2}
\int_{x_F}^{1} dx_p \, {x_p\over x_F} \Bigg[
f_{q/p} (x_p,Q_f^2)~ N_F \left({x_p\over x_F} p_t , b\right)~ D_{h/q} 
\left({x_F\over  x_p}, Q_f^2\right) +  \nonumber \\
&&
f_{g/p} (x_p,Q_f^2)~ N_A \left({x_p\over x_F} p_t , b\right) ~ 
D_{h/g} \left({x_F\over x_p}, Q_f^2\right)\Bigg]~,
\label{eq:final}
\end{eqnarray}
where $p_t$ and $x_F$ are the transverse momentum and the Feynman-$x$
of the produced hadron, respectively.  $x_p$ denotes the momentum
fraction of a projectile parton and $b$ is the impact parameter.  Note
that this expression is different from the more common $k_t$
factorized expressions~\cite{bfkl,kt_factor} in that {\it the
radiation vertex is exact}. The common eikonal (soft gluon)
approximation made in the $k_t$-factorization approach is
omitted, which reflects in the fact that the projectile parton
distribution functions $f(x_p,Q_f^2)$, and their fragmentation
functions $D(z, Q_f^2)$ into hadrons, evolve according to the full
DGLAP~\cite{dglap} evolution equations and respect the momentum
sum-rule.  The importance of this evolution for the interpretation of
forward-rapidity data from RHIC is discussed in detail in~\cite{aaj}.

Eq.~(\ref{eq:final}) resums large logarithms of $1/x_A$ arising from
evolution of the target wave function with energy as well as
logarithms of $Q^2$ from evolution of the collinearly factorized
projectile quark and gluon distributions with $Q^2$.  It does not
include large-$x$ effects such as recoil in the nucleus, nor does it
account for logarithms of $Q^2$ on the target side (only
collinear emissions from the projectile are included, c.f.\
ref.~\cite{aaj} for details).  Therefore, this expression is valid
when the projectile partons are in either the full or double-log DGLAP
kinematics (in other words, above the ``extended'' geometric scaling
window for the projectile) while the target is in the small-$x_A$,
high-density regime.

To illustrate the regime of applicability of eq.~(\ref{eq:final}) from
another perspective we compare to the usual $k_t$-factorization approach,
where the single-inclusive gluon production cross section is given by
\be \label{kt-factorized}
\frac{d\sigma}{dy d^2p_t} = F
       \int dk_t^2 \, \frac{p_t^2}{k_t^2 (k_t-p_t)^2} \,
       \phi_p(x_p,k_t^2)\, \phi_A(x_A,(k_t-p_t)^2)~.
\ee
The pre-factor $F=4\pi\alpha_s N_c/(N_c^2-1)/p_t^4$ and
$\phi_{p,A}(x_{p,A},k_t^2)=\partial x_{p,A}g(x_{p,A},k_t^2)/(\partial \log k_t^2)$ 
denote the unintegrated gluon distribution functions of
the proton and of the nucleus, respectively.

To leading $\log p_t^2$ accuracy we only need to
pick up the contributions from $k_t\sim0$ and $k_t\sim p_t$:
\be
\frac{1}{F} \frac{d\sigma}{dy d^2p_t} &\approx&
       \int_0^{p_t^2} dk_t^2 \frac{1}{k_t^2}
       \phi_p(x_p,k_t^2) \phi_A(x_A,p_t^2)  +
       \int_0^{p_t^2} dk_t^2 \frac{1}{k_t^2}
       \phi_p(x_p,p_t^2) \phi_A(x_A,k_t^2)  \label{eq3} \\
 &=&
       x_p\, g(x_p,p_t^2)\, \phi_A(x_A,p_t^2) +
       \phi_p(x_p,p_t^2)
       \int_0^{p_t^2} dk_t^2 \frac{1}{k_t^2} \phi_A(x_A,k_t^2)~. \label{eq4}
\ee
In the second term, we shifted $(k_t-p_t)^2\to k_t^2$.

The first term corresponds to eq.~(\ref{eq:final}), if one replaces
the $D_{h/g}(z)$ fragmentation function from there by $\delta(1-z)$ (we
are now discussing gluon, not hadron production) and identifies
\be
F \, \phi_A(p_t) = \frac{N_{A}(p_t)}{(2\pi)^2}~.
\ee
Nevertheless, one should keep in mind that eq.~(\ref{kt-factorized})
is derived in the recoilless approximation and so is not applicable
when $x_p$ is large, say $\ge0.01$ for a proton projectile. As already
mentioned above, eq.~(\ref{eq:final}) was derived without this
limitation~\cite{aaj}. 

On the other hand, the second term from~(\ref{eq4}) is clearly not
part of eq.~(\ref{eq:final}). For $p_t\sim Q_{gs}$ (the geometric
scaling momentum of the nucleus) the unintegrated gluon  distribution
functions are given by (see e.g.~\cite{KLM})
\be
\phi_p(x_p,p_t^2) \sim 1~~~,~~~ \phi_A(x_A,k_t^2)\sim S_A\,k_t^2\left(
\frac{Q_s^2(x_A)}{k_t^2}\right)^\gamma~.
\ee
Here, $S_A$ denotes the transverse area of the nucleus and $\gamma<1$ is
the anomalous dimension of its gluon distribution function, see
below. Constant prefactors which are irrelevant for the present
discussion have been dropped. Using these expressions
in~(\ref{eq3}) shows that no logarithms of $p_t^2$ arise
from the second term, which therefore is suppressed. On the other
hand, for $p_t\gg Q_{gs}(x_A)$, the integral over $k_t^2$ in the
second term of~(\ref{eq4}) {\em does} give rise to a contribution
$\sim\log(p_t^2/Q^2_{gs})$ since $\phi_A(x_A,k_t^2)\sim1$ for
$k_t>Q_{gs}$. Hence, for $p_t\gg Q_{gs}$ collinear emissions from the
target have to be included as well, for example by using CCFM
unintegrated gluon distributions~\cite{CCFM}.

The purpose of the present manuscript is to check for the presence of
geometric scaling violations in the mid-rapidity data from
RHIC. Consequently, we focus on the regime of moderately large transverse
momenta not too far above $Q_{gs}$. Since there $x_p\gsim0.01$, we believe
that it is more appropriate to employ eq.~(\ref{eq:final}) rather
than~(\ref{kt-factorized}).

The cross section depends on the scattering probability of dipoles of
size $r_t$ in the fundamental and adjoint representations,
respectively, at an impact parameter $b$:
\be 
N_F (r_t,b) &\equiv & {1\over N_c} \, {\rm Tr_c} \,
\langle 1- V^{\dagger}(b - r_t/2) V(b + r_t/2)\rangle, \nonumber \\ 
N_A(r_t,b) &\equiv & {1\over N^2_c -1} \, {\rm Tr_c} \;
\langle 1-U^{\dagger}(b -r_t/2) U(b + r_t/2)\rangle~,
\label{eq:N_FA}
\ee
where $V$ and $U$ denote Wilson lines along the light
cone~\cite{jimwlk} in the corresponding
representation, and $N_c$ is the number of colors. 

In practice, the ab-initio determination of dipole profiles by
solution of the JIMWLK equations is not feasible yet. Therefore, one
resorts to theory-motivated phenomenological parametrizations of the
dipole profile which can be tested against data.  An example is
the KKT model~\cite{dima}, where the dipole profile is parameterized as
\be  \label{NA_param}
N_A(r_t,y_h) = 1-\exp\left[ - \frac{1}{4} (r_t^2 Q_s^2(y_h))^
{\gamma(y_h,r_t)}\right]~,
\ee
with $y_h$ the rapidity of the produced hadron.
The saturation scale of the nucleus at $y_h$ is given
as\footnote{Note that if hadron masses and intrinsic transverse
  momenta from fragmentation are neglected, then the rapidity of the
  produced hadron equals that of its parent parton. Hence, we do not
  distinguish between them.}
\be 
\label{Qs_y}
Q_s(y_h) = Q_0 \exp [\lambda (y_h-y_0)/2]~.
\ee
Here, $\lambda\simeq 0.3$ is fixed by the DIS data 
and the initial condition $Q_0\simeq1$~GeV is set at $y_0=0.6$,
where small-$x$ quantum evolution becomes effective.
The saturation scale  for the 
dipole in the fundamental representation differs by a factor of  
$Q_s^2\to Q_s^2\, C_F/C_A = \frac{4}{9} Q_s^2$. 

$\gamma$ denotes the anomalous dimension of the target gluon distribution
with saturation boundary condition and is modeled by KKT as
\be
\gamma(y_h,q_t) = \frac{1}{2}\left(1+
\frac{|\xi(y_h,q_t)|}{|\xi(y_h,q_t)|+\sqrt{2|\xi(y_h,q_t)|}
+7\,c\,\zeta(3)}\right)~.
\label{eq:gam}
\ee
Here, we have written $\gamma$ as a function of transverse momentum
rather than dipole size by evaluating it at a characteristic value
$r_t \sim 1/q_t$. The free parameter $c$ in~(\ref{eq:gam}) governs
the onset of the quantum evolution regime, and has been fixed
by KKT to $c=4$.

The {\em Ansatz}~(\ref{NA_param}) exhibits geometric scaling when
$\gamma$ is constant. Scaling violations are introduced into the KKT
parameterization through the function
\be
\xi(y_h,q_t) = \frac{\log (q_t^2/Q_0^2)}{(\lambda/2)(y_h-y_0)}~.
\label{eq:xi}
\ee
$\xi$ vanishes as $y_h\to\infty$ at fixed $q_t$, so that $\gamma\to 1/2$.
On the other hand, approaching the saturation boundary at large
rapidity far from the initial condition (i.e.\ $q_t\to Q_s(y_h)$ for
$y_h\gg y_0$) leads to $\xi\to2$. Due to the last term in the
denominator of eq.~(\ref{eq:gam}) this corresponds to
$\gamma\approx0.53$, again close to the usual BFKL saddle point {\em
  without} saturation boundary conditions.


\section{Scaling violations in the mid-rapidity RHIC data}
\label{sec_ScalViol}

It was shown in ref.~\cite{aaj} that the BRAHMS
data~\cite{Arsene:2004ux} on forward hadron production from d+Au
collisions at RHIC are described very well\footnote{Already at leading
order in $\alpha_s$, if a phenomenological $p_t$-independent
$K$-factor for NLO corrections is allowed for.} by both the
KKT~\cite{dima} and the IIM~\cite{IIM} dipoles, provided that DGLAP
evolution of the dilute projectile with exact splitting functions is
taken into account.  The predictions made in \cite{aaj} appear to be 
confirmed by the preliminary data from STAR on $\pi^0$ production at rapidity
$y_h\simeq4$~\cite{greg}. The fact that both dipole parameterizations
generically include an anomalous dimension of the target gluon
distribution that {\em differs} from 1 at small $x_A$ proved
crucial.  In particular, for $y_h\gsim3$ both KKT and IIM dipoles predict
a nearly $p_t$-independent anomalous dimension $\gamma\approx 0.6$ for
$p_t\lsim5$~GeV, which is close to the LO-BFKL prediction\footnote{This
  is the anomalous dimension for BFKL evolution with saturation
  boundary condition, i.e.\ for evolution along the saturation line.}
of $\gamma=\gamma_s\approx 
0.627$ and, with eq.~(\ref{eq:final}) from above, is consistent with
the experimental $p_t$ distributions~\cite{aaj}.

In the present paper, we consider particle production in the central
region, $y_h=0$. As we shall see, here the behavior of $\gamma$
required by the data and the CGC approach is very different. Before
comparing to data, however, it is useful to analyze the kinematics.
While forward hadron production probes rather small $x_A$ in the
target nucleus (c.f.\ appendix~B in ref.~\cite{aaj}), one may wonder
to what extent the conditions for applicability of the CGC formalism
are met in the central rapidity region, for RHIC energy ($\sqrt
s=200$~GeV). We expect that the CGC description breaks down beyond
some $x_0$, where recoil effects affect the evolution with
$\log\,1/x$. Near $x_0$, knowledge of the {\em width} $\Delta x/x_0$
of that transition region would also be required.  Neither $x_0$ nor
$\Delta x$ can be determined from the CGC formalism itself, since we
presently lack a treatment of recoil and of other ``large-$x$''
effects within this approach.

However, since evolution in $x$ appears to be quite rapid, it is
probably not a bad approximation to consider a sharp boundary $x_0$
beyond which the CGC small-$x$ formalism will break down.
Experimentally, it is known that HERA data on proton structure
functions at or below $x_0 \sim 0.01$ can be described reasonably well
within the small-$x$ evolution
approximation~\cite{GBW,Stasto:2000er,IIM,Forshaw}.  In case of
nuclear targets, larger values of $x_0$ are expected due to the $\sim
A^{1/3}$ enhancement of the saturation scale near
$x_0$~\cite{MV}. Therefore, for a rough estimate, we take $x_{0} =
0.05$ as the upper limit of validity of the small-$x$ approximation
for the target.

At $y_h=0$, the projectile and target $x$ are equal and given by
$x=q_t/\sqrt{s}$, with $q_t$ the transverse momentum of the produced
parton. If we assume that the small-$x$ approximation of the CGC can
be applied for $x<x_0 \simeq 0.05$, then this translates into
transverse momenta up to $q_t \simeq 0.05\sqrt s$. Thus, for RHIC
energy the approach should be valid for $q_t \lsim 10$~GeV.
Hadronization reduces this number by a factor of $\approx0.7$,
according to the typical values
of $z=x_F/x_p$ in the fragmentation. Hence, we estimate that
in the central region the small-$x$ regime extends roughly
up to hadron transverse momenta of $p_t \simeq 7$~GeV.

\begin{figure}[hbt]
\centering
\centerline{\epsfig{figure=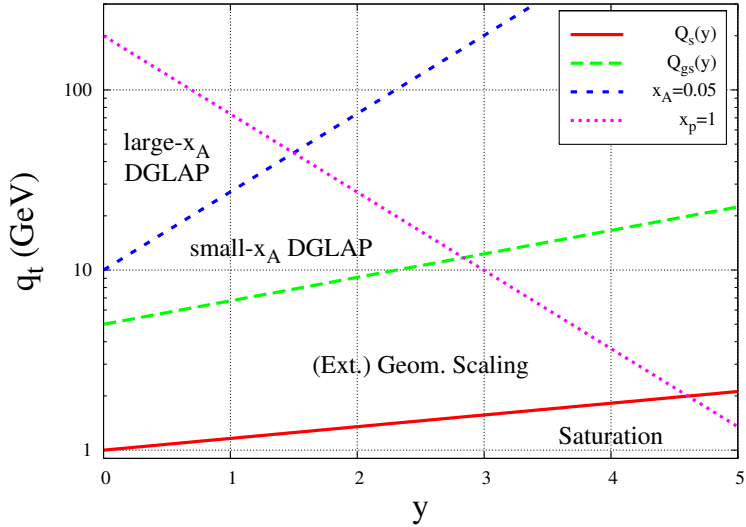,width=4in}}
\caption{Schematic ``phase diagram" of pA collisions at RHIC. For this
plot, we have simply assumed that $Q_s(y)=1~$GeV$\times\exp(\lambda y/2)$
and $Q_{gs}(y)=Q_s^2(y)/200$~MeV; see, however, sections~\ref{sec_NewDipole}
and \ref{sec_applic}.}
\label{fig:gs_window}
\end{figure}
This upper limit for the small-$x$ approximation is above the boundary
between the extended scaling and the dilute DGLAP regimes,
$Q_{gs}(y)\sim Q_s^2(y)/\Lambda$~\cite{ScalingViol}. With
$Q_s^2\simeq1$~GeV$^2$ at mid-rapidity (averaged over the transverse
plane) and $\Lambda=200$~MeV, this boundary occurs at a parton
transverse momentum of about $q_t\simeq5$~GeV (or hadron transverse
momentum $p_t\simeq3.5$~GeV). By this $q_t$, the anomalous dimension
of the target gluon distribution should already have approached its
DGLAP limit, starting from $\gamma=\gamma_s$ at $q_t = Q_s \simeq
1$~GeV. Hence, we expect large scaling violations in the central
region of p+A collisions at RHIC already for transverse momenta on the
order of a few GeV, signaling the approach towards the DGLAP
regime. These different kinematical regimes are
summarized in Figure~\ref{fig:gs_window}. According to the discussion
in the previous section, our approach via eq.~(\ref{eq:final}) is
valid up to $q_t\simeq Q_{gs}$ but not much beyond.

These estimates are in qualitative agreement with the results of
Accardi and Gyulassy~\cite{AccGy}. They find that a resummation of the
Glauber multiple-scattering series (which basically corresponds to
semi-classical saturation) is required at $p_t\sim1$~GeV, and that a
rapid transition to the DGLAP regime occurs towards higher transverse
momenta. Comparing their approach to data from RHIC, they do not seem
to find evidence for a broad geometric scaling window.  In this vein,
we also recall that a purely semi-classical model without quantum
evolution does not reproduce the RHIC mid-rapidity data, as shown
in~\cite{Baier:2005dz}\footnote{A fit is possible if one assumes an
ad-hoc large ``non-perturbative'' contribution to $Q_s$, which has to
be dropped again in the forward region. A critical discussion of this
issue is given in ref.~\cite{Baier:2005dz}.} (for an alternative approach, 
see \cite{Qiu:2004da}).

Similarly, leading-twist NLO pQCD~\cite{werner} provides an accurate
description of the mid-rapidity cross-section for $p_t\gsim5$~GeV. We
point out that while this indeed contradicts the original GBW model,
it is perfectly consistent with improved HERA fits including scaling
violations~\cite{IIM} and with the CGC theory, which does predict that
$\gamma\to\,\sim1$ for $q_t> Q_{gs}$.  That this is perhaps the
correct interpretation can be inferred from the {\em suppression} of
scaling violations at large rapidity (and $p_t\sim$ few GeV), which
again is a generic prediction of the CGC approach\footnote{Deriving
from the fact that $\Delta\gamma\sim1/y$ at fixed $p_t$. See also
refs.~\cite{CroninEvol}.}, and has already been verified to be
consistent with RHIC data~\cite{aaj}.

\begin{figure}[hbt]
\centering
\centerline{\epsfig{figure=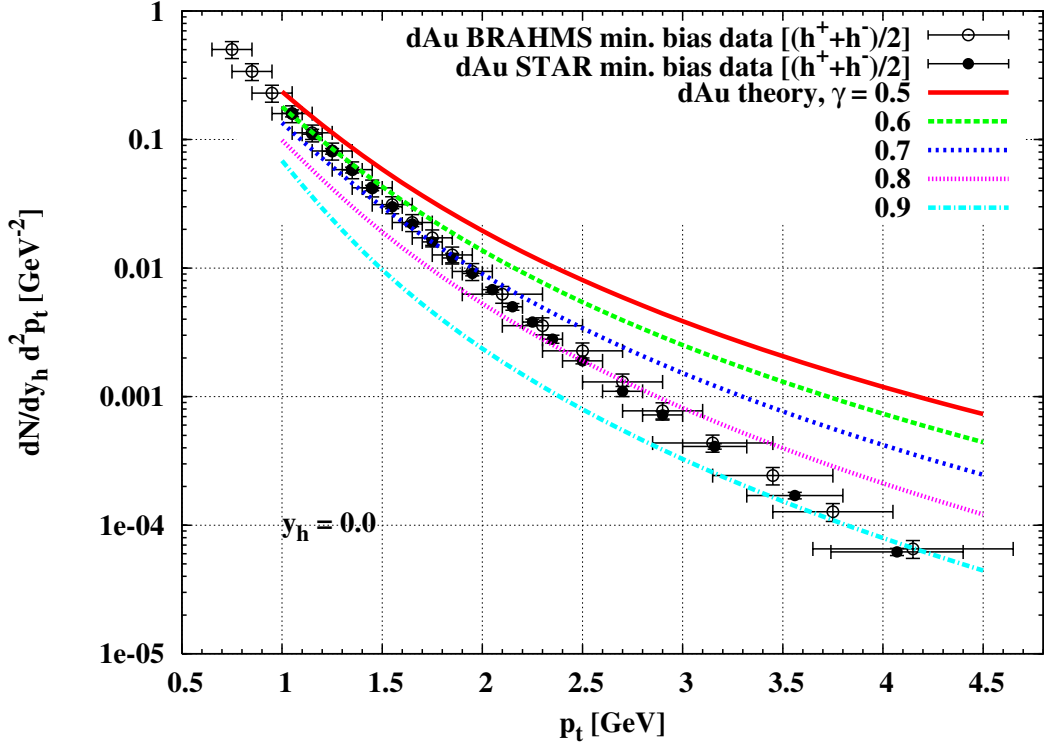,width=4in,angle=-90}}
\caption{Charged hadron $p_t$ spectrum obtained from
  eq.~(\ref{eq:final}) using the KKT dipole for various fixed
  anomalous dimensions $\gamma$. The points with error bars show the
  BRAHMS~\cite{Arsene:2004ux} and STAR~\cite{STAR_y0} data for
  minimum-bias d+Au collisions (RHIC, mid-rapidity).}
\label{fig:mid_rhic}
\end{figure}
In Fig.~\ref{fig:mid_rhic} we show the transverse momentum spectra of
charged hadrons at mid-rapidity for minimum-bias d+Au collisions at
RHIC\footnote{For better visibility, we show only the
BRAHMS~\cite{Arsene:2004ux} and STAR~\cite{STAR_y0} measurements. Data at
central rapidity has also been obtained by the PHENIX~\cite{PHENIX_y0}
collaboration.}. We also show the theory curves obtained from
eq.~(\ref{eq:final}), using CTEQ5-LO~\cite{cteq} parton distribution
functions for the projectile deuteron, and KKP-LO~\cite{KKP}
fragmentation functions at the scale $Q_f = p_t$. We plot curves for
various {\em fixed} anomalous dimensions from $\gamma=0.5$ to $0.9$.
Clearly, a rather steep rise of the anomalous dimension in the $p_t
\sim 1 - 4$ GeV range is required for a satisfactory description of
the mid-rapidity data. At the lower end, $p_t\approx1$~GeV, the
anomalous dimension is close to $0.6$, perhaps marking the onset of
saturation dynamics in the weak-coupling regime. This observation is
further supported by the fact that saturation based models
successfully describe the hadron multiplicities at RHIC, which are
dominated by soft hadrons~\cite{kt_factor}.  On the other hand, at
$p_t=4$~GeV, the anomalous dimension $\gamma\approx0.9$ required to
fit the data nearly equals its leading twist DGLAP limit of 1.

\begin{figure}[hbt]
\centering
\centerline{\epsfig{figure=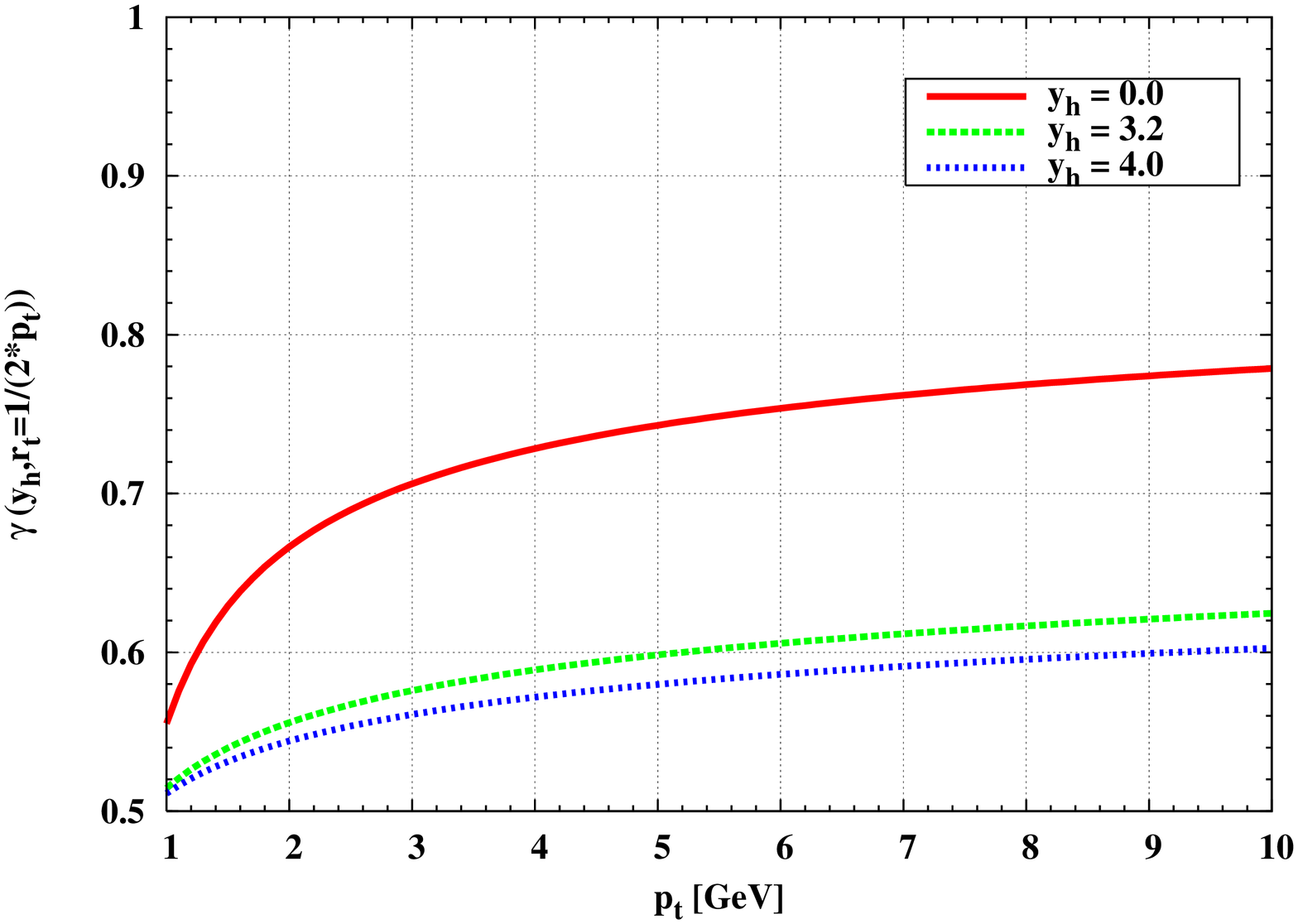,width=3in,angle=-90}}
\caption{Anomalous dimension $\gamma(y_h,r_t)$ of the small-$x$
  nuclear gluon distribution for the KKT dipole model, evaluated at
  $r_t=1/(2p_t)$.}
\label{fig:KKT_gamma_ptdep.ps}
\end{figure}
To illustrate this further, we plot 
the transverse momentum dependence of $\gamma$ for
the KKT model in Fig.~\ref{fig:KKT_gamma_ptdep.ps}. For large
transverse momenta, we assume that the Fourier transform of $N_A(r_t)$
is dominated by transverse distances of order $1/q_t\sim1/(2p_t)$ and
hence evaluate $\gamma(r_t)$ at $r_t=1/(2p_t)$.

One observes that the KKT parameterization of the anomalous dimension
does not vary much with transverse momentum, staying within
$\approx20\%$ of $\gamma\approx0.6$.  As discussed above, this
behavior is generic to the CGC at {\em large rapidity} (i.e.\ small
$x_A$), and is consistent with the RHIC data~\cite{aaj}. However, we
have also seen that the data requires a rapid approach towards the
DGLAP regime at mid-rapidity and $p_t\sim4$~GeV, while
$\gamma_{KKT}\to1$ only at asymptotically high $p_t$.  In other words,
we find evidence for substantial scaling violations at mid-rapidity,
larger than those incorporated into the KKT parameterization but in
agreement with our estimates on $Q_{gs}$ from above, which then weaken
at large rapidity (they get pushed to higher $p_t$).  Hence, the
qualitative behavior of $\gamma$ implied by the RHIC data is
remarkably consistent both with HERA phenomenology and with the
predictions of the CGC approach.

\section{New parameterization of the anomalous dimension}
\label{sec_NewDipole}

Motivated by these results, we introduce a new parameterization of the
dipole profile with an anomalous dimension incorporating stronger
scaling violations than the KKT dipole. At the same time, it corrects
some deficiencies of the IIM dipole at large $Q^2$.  In our
parameterization, the anomalous dimension is given by 
\be
\gamma(Q^2,Y) &=& \gamma_s + \Delta\gamma(Q^2,Y) \nonumber\\
\Delta\gamma &=& (1-\gamma_s)\,\frac{\log (Q^2/Q_s^2)}{\lambda\, Y
+\log(Q^2/Q_s^2) + d\sqrt{Y}}~.  \label{eq:gam_new} 
\ee 
We have written $\gamma$ in terms of the variables appropriate for
DIS, i.e.\ $Q^2$ and $Y=\log(1/x)$, similar to the GBW and IIM
models. Here, $Q^2\equiv1/r_t^2$ is the {\it inverse transverse
size of the dipole}, not to be confused with the factorization scale
which enters the parton distribution and fragmentation functions in
eq.~(\ref{eq:final}).  We shall transform into the variables
appropriate for hadron-hadron collisions shortly.

The function is constructed such as to satisfy the following limits:
\begin{enumerate}
\item at any fixed rapidity $Y$, $\gamma\to1$ for $Q^2 \to\infty$.
\item if $Y\to\infty$ at fixed $Q^2/Q_s^2(Y)$, $\Delta\gamma$ decreases
  asymptotically like $\sim 1/Y$, as determined by
  BFKL~\cite{IIM}. This indicates that the geometric scaling window
  broadens with $Y$. 
\item $\gamma\to\gamma_s$ for $Q^2\to Q_s^2(Y)$ at any rapidity.
\item At large but fixed rapidity, $Y\gg (d/\lambda)^2$, the geometric
  scaling window reaches up to $\log Q^2_{gs}(Y)/Q_s^2(Y)\sim\lambda Y$,
  consistent with the asymptotic estimate for
  the geometric scaling window~\cite{ScalingViol}: 
$Q^2_{gs}\sim Q_s^4/\Lambda^2_{QCD}$.
\end{enumerate}
This last condition is the most essential difference between our new 
parameterization and the IIM parameterization~\cite{IIM}, where
\be
\gamma_{IIM}(Q^2,Y) &=& \gamma_s +
\frac{\log (Q^2/Q_s^2)}{\kappa\, \lambda\, Y}~.
\label{eq:gam_IIM}
\ee
Here, one finds $\log Q^2_{gs}/Q_s^2\sim (1-\gamma_s)\kappa
\lambda Y$, with $\kappa\simeq 10$. Hence, at large $Y$, their
$Q^2_{gs}(Y)$ is much larger than $Q_s^4(Y)/\Lambda^2$.
Such behavior arises if the geometric scaling window is estimated from
the diffusion term in the expansion of the LO-BFKL solution to second
order about the saturation saddle-point. On the other hand, if
$Q_{gs}^2$ is estimated via the transition point between the LLA and DLA
saddle points, respectively, one finds $\log
Q^2_{gs}/Q_s^2\sim\lambda Y$, without the additional factor of
$\kappa$ (for more details see, for example, section~2.4.3 in
ref.~\cite{JK} and references therein).

With $\kappa\lambda Y$ replaced by $\lambda Y$, however,
$\Delta\gamma$ rises too rapidly with $\log Q^2$ to allow for a good
description of the RHIC data. Also, one should keep in
mind that the IIM parameterization in fact provides a good fit to the
HERA data on $F_2(x,Q^2)$ at small $x$. Hence, strong modifications of
$\gamma$ are not desirable.  As a consequence, to restore
(approximate) agreement with the IIM parameterization, we are led to
introduce terms which are subleading in $Y$. This agrees with indications
from studies of small-$x$ evolution equations which show that at
realistic energies (rapidities), subleading corrections to the
asymptotic expressions are quite large. In~(\ref{eq:gam_new}), these
are modeled by the $\sqrt{Y}$ term with a free coefficient. From the
mid-rapidity RHIC data, we extract $d\simeq1.2$ (see next section).

\begin{figure}[hbt]
\centering
\centerline{\epsfig{figure=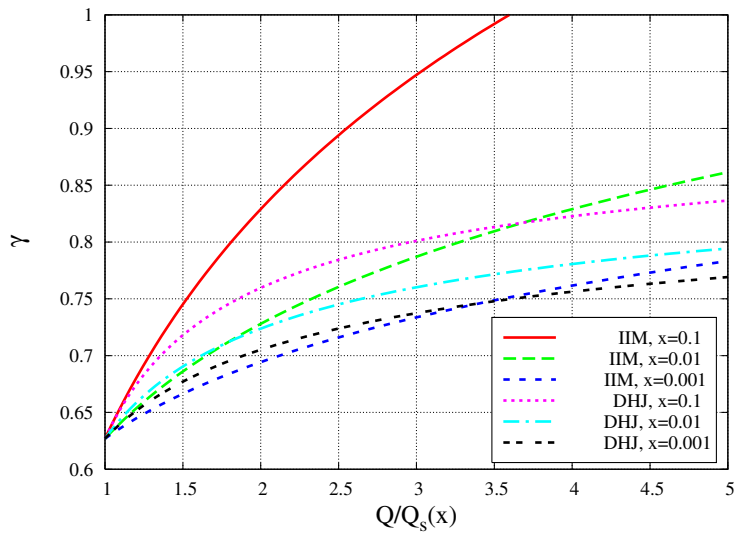,width=4in}}
\caption{Anomalous dimension versus $Q$ at various $x$.}
\label{fig:gamma_IIM_DHJ}
\end{figure}
Fig.~\ref{fig:gamma_IIM_DHJ} compares our dipole
parameterization~(\ref{eq:gam_new}) with $d=1.2$ to that from
eq.~(\ref{eq:gam_IIM}). At small $x$ they are very similar as $\gamma$
becomes flatter and approaches $\gamma_s$. Towards large $x$, $\gamma$
is somewhat smaller for our parameterization\footnote{Ref.~\cite{IIM}
considered only the region $x<0.01$ and the fact that $\gamma_{IIM}$ is not
bounded from above was irrelevant for their analysis.}. At the
same time, the new anomalous dimension is closer to 1 than $\gamma_{KKT}$, 
which was shown in Fig.~\ref{fig:KKT_gamma_ptdep.ps}~\footnote{Note also the
  factor of 2 difference in the $p_t$-scale at which $\gamma$ is
  evaluated.}, i.e.\ our parameterization features a narrower scaling
window. With $d=1.2$, which restores agreement with the IIM dipole at
$x=10^{-3}$ and also describes the RHIC d+Au data correctly,
the subleading term in the rapidity evolution of $\gamma$
actually dominates until $Y\simeq (d/\lambda)^2=16$.

Finally, we also point to another recent parameterization of the
dipole profile in ref.~\cite{Baier:2005dz}, which is meant to
reproduce the RHIC data. That parameterization also includes breaking
of geometric scaling at mid-rapidity, but not the sub-asymptotic terms
for the evolution with rapidity. The dipole profile at large rapidity
again resembles the KKT, IIM, and our present dipole model and so the
forward hadron spectra ``generated'' by all these parameterizations
should be rather similar.

\section{Application to d+Au collisions at RHIC}
\label{sec_applic}
For hadron-hadron collisions, we need to replace $Q^2\to1/r_t^2$ and
Fourier transform the dipole profile function from~(\ref{NA_param}):
\be   \label{FTdipole}
N_{A,F}(q_t) = \int d^2r_t \;e^{i\vec{q}_t\cdot\vec{r}_t}N_{A,F}(r_t)=
2\pi\int_0^\infty dr_t \;r_t \;J_0(r_t \,q_t)\;N_{A,F}(r_t)~.  
\ee 
Unfortunately, it is very challenging to
perform this Fourier transform numerically when $\gamma$ is not
constant but $r_t$-dependent. It is therefore common to replace
$\gamma(r_t)$ by a constant $\gamma(r_t=1/q_t)$, which makes the
Fourier transform tractable~\cite{dima,aaj}. This is a reasonable
approximation in the region where $\gamma(r_t)$ is rather flat,
e.g.\ at large rapidity and $p_t \ll Q_{gs}$. On the other hand, 
at mid-rapidity, where scaling violations are larger, we found that 
it is important to account for the $r_t$-dependence of the anomalous 
dimension, as this strongly affects $N(q_t)$.

However, with a $r_t$-dependent anomalous dimension, we were unable to
perform the Fourier transform~(\ref{FTdipole}) numerically at very
large transverse momentum, where the phase factor oscillates extremely
rapidly.  More importantly, the Fourier transform turns negative
already at intermediate $q_t$ ($\approx 8$~GeV for gluons at
mid-rapidity), where our numerical methods are still reliable. This unphysical
behavior occurs for all of the above-mentioned $\gamma(r_t)$
parameterizations (KKT, IIM and ours) and presumably indicates that
the behavior of the function at very small $r_t$ is 
inadequate~\footnote{This problem was encountered also in earlier 
studies \cite{fgjjm} of Fourier transforms of
the Bartels {\it et al.} DGLAP-improved dipole profile \cite{GBW}.}. We are
nevertheless able to determine the transverse momentum spectrum of
hadrons reliably up to about $p_t=4$~GeV, which doesn't receive
contributions from the unphysical oscillations of $N_A(q_t)$ at high
parton momentum. As discussed in sec.~\ref{sec2}, we do not expect
that eq.~(\ref{eq:final}) is valid much beyond this $p_t$ 
at RHIC (mid-rapidity) anyways.

The field of the target nucleus is probed at the rapidity $y_A$ given
by $y_A=\log 1/x_A=\log(1/x_p)+2y_h$ (see appendix B in~\cite{aaj} for
details of kinematic definitions). Hence, the saturation scale is
defined at that rapidity, 
\be
Q_s^2(x_A)= Q_0^2\, A_{\rm eff}^{1/3}\, (x_0/x_A)^{\lambda}~.
\ee
This implies that at fixed hadron rapidity $y_h$, as we sum the
contributions from projectile partons with different momentum
fractions $x_p$ in eq.~(\ref{eq:final}), $Q_s(x_A)$ also varies.

The effective mass number of the nucleus depends
on the impact parameter; following ref.~\cite{dima}, for heavy
$A\sim200$ targets we do not perform the integral over $b$ explicitly
but take $A_{\rm eff}=18.5$ for minimum bias collisions (the issue of
impact parameter averaging is also discussed in
ref.~\cite{Baier:2005dz}).  We fix the scale $x_0=3\cdot10^{-4}$ to the
value extracted in~\cite{GBW} by a fit to HERA data.  With $Q_0 =
1$~GeV, one obtains $Q_s\simeq1$~GeV at mid-rapidity for transverse
momenta $\sim 1$~GeV, which is reasonable~\cite{dima}.

\begin{figure}[hbt]
\centering
\centerline{\epsfig{figure=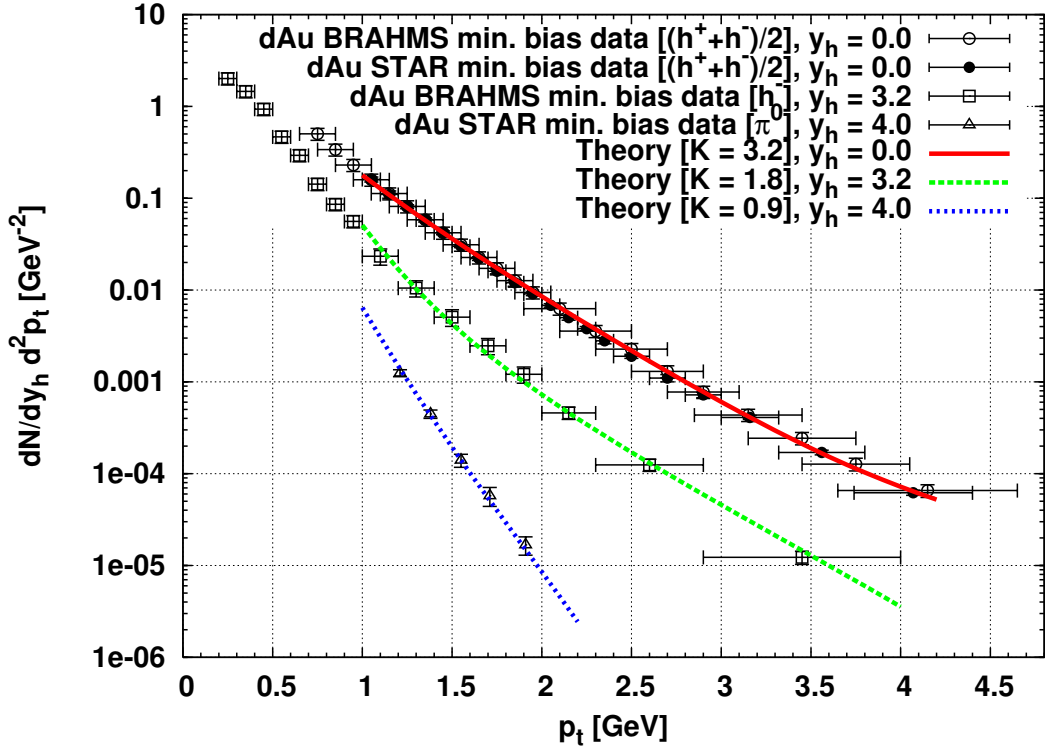,width=4in,angle=-90}}
\caption{Comparison of theory and
  data~\protect\cite{Arsene:2004ux,greg,STAR_y0} for minimum-bias d+Au
  collisions at RHIC energy.}
\label{fig:dA_new}
\end{figure}
In Fig.~\ref{fig:dA_new} we show our results for minimum-bias d+Au
collisions at RHIC energy with the new
parameterization~(\ref{eq:gam_new}). The factorization scale in
eq.~(\ref{eq:final}) is taken as $Q_f = p_t$ for all rapidities.  All
other parameters have been fixed above. As promised, the shape of the
mid-rapidity $p_t$-distribution is described quite well. For the
correct overall normalization, we need to multiply our LO calculation
by a factor $K_{y=0}\simeq3$. This is much larger than at forward
rapidity, were $K_{y=3}\simeq2$ and $K_{y=4}\simeq 1$, respectively
(for the same scale $Q_f = p_t$ and the same parton
distribution~\cite{cteq} and fragmentation~\cite{KKP} functions). Such
a rapid increase of the $K$-factor towards mid-rapidity is expected
because of larger phase space for NLO corrections~\footnote{This trend
is also present in pQCD calculations of hadron spectra in
proton-proton collisions~\cite{wv}.}. Despite the good agreement of
the LO curve with the $p_t$-dependence of the data, NLO calculations
of particle production are clearly desirable.

Fig.~\ref{fig:dA_new} also shows that the new parameterization of the
anomalous dimension introduced here describes the forward data as
well as the IIM and KKT dipole models (for the latter, see curves
in~\cite{aaj}). As already alluded to above, the reason for this
generic behavior is that all models predict a rather flat
$\gamma(r_t)\approx\gamma_s$ at small $x$ (wide geometric scaling
window), and hence do not differ much. This also reflects in
the fact that the forward rapidity curves from Fig.~\ref{fig:dA_new},
which were obtained with the $r_t$-dependent anomalous dimension, are
quite similar to those shown in ref.~\cite{aaj}, which employed a constant
$\gamma= \gamma(r_t=1/2p_t)$. We stress that such good agreement with
the data in the forward region is only achieved if the projectile
parton distributions satisfy DGLAP evolution with {\em exact} rather
than small-$x$ approximated splitting functions~\cite{aaj}.


\section{Summary} \label{sec_summary}

In summary, in this paper we extended our previous studies of forward
rapidity hadron production in deuteron-gold collisions at RHIC to the
central rapidity region. We argued that here the CGC predicts
significantly larger scaling violations than at forward rapidity, and
that indeed a relatively rapid transition of the anomalous dimension
of the target gluon distribution from $\gamma_s\approx0.63$ (BFKL with
saturation boundary conditions) to $\gamma\to1$ (DGLAP) is required by
the data.  It is important to note that this transition occurs in a
regime where the small-$x$ approximation made in the Color Glass
Condensate formalism is still valid. Hence, the scaling violations
predicted by the CGC can indeed be tested.

We propose a new parameterization of the dipole profile which features
a steeper transverse momentum dependence than the KKT dipole and leads
to a satisfactory description of the mid-rapidity data. This agreement
supports the predictions of the Color Glass Condensate formalism regarding
the existence of different kinematic regions with different underlying
physics.

Schematically, at small transverse momentum one probes the saturation
region of the nucleus, where the underlying physics is that of a
``black'' target. As the transverse momentum increases, there is a
transition from the saturation to the (extended) geometric scaling
regime, corresponding to an anomalous dimension close to that from
BFKL with saturation boundary condition. A further increase in the
transverse momentum takes one beyond the geometric scaling region,
signaled by the necessity of introducing rather large scaling breaking
in the dipole profile. We showed that for transverse momenta of about
$4$~GeV, one is already in a small-$x$ DGLAP regime. Thus, the
transition from the saturation to the DGLAP regions is quite narrow in
the central rapidity region at RHIC. On the other hand, at large
rapidity ($y_h\gsim3$) only the saturation and scaling regions remain,
while the DGLAP window is cut off by finite-energy constraints, which
has not been realized before. At LHC energy, all these kinematic
regimes extend to even higher $p_t$, providing a great opportunity to
study the various ``phases'' of high-energy QCD in more detail and
with better accuracy.

\section*{Acknowledgements}

We would like to thank W.\ Vogelsang for helpful discussions. 
J.J-M.\ is supported in part by the U.S.\ Department of Energy under 
Grant No.\ DE-FG02-00ER41132.

\end{document}